\documentclass[prl,twocolumn,floatfix,showpacs,superscriptaddress]{revtex4}
\usepackage{graphics}
\usepackage{epsfig}
\usepackage{times}
\usepackage{bm}
\usepackage{color}

\renewcommand{\i}{{\rm i}}

\usepackage{amsmath}	% required for `\align' (yatex added)
\begin{document}
%\draft
\title{Metal-insulator transition of the Kagom\'{e} lattice fermions at
1/3 filling}
\author{Satoshi Nishimoto}
\affiliation{Leibniz-Institut f\"ur Festk\"orper- und Werkstoffforschung 
Dresden, D-01171 Dresden, Germany}
\author{Masaaki Nakamura}
\affiliation{
Department of Physics, Tokyo Institute of Technology,
%Oh-Okayama, Meguro-ku, 
Tokyo 152-8551, Japan}
\author{Aroon O'Brien}
\affiliation{
Max-Planck-Institut f\"{u}r Physik komplexer Systeme,
N\"{o}thnitzer Stra{\ss}e 38, D-01187 Dresden, Germany}
\author{Peter Fulde}
\affiliation{
Max-Planck-Institut f\"{u}r Physik komplexer Systeme,
N\"{o}thnitzer Stra{\ss}e 38, D-01187 Dresden, Germany}
\affiliation{Asia Pacific Center for Theoretical Physics, Pohang, Korea}
\date{\today}
%%{submitted in }
\begin{abstract}
We discuss the metal-insulator transition of the spinless fermion model
on the Kagom\'{e} lattice at 1/3-filling. The system is analyzed using
exact diagonalization, the density-matrix renormalization group methods
and the random phase approximation. In the strong-coupling region, the
charge-ordered ground state is consistent with the predictions of the
effective model with a plaquette order. We find that the qualitative
properties of the metal-insulator transition are totally different
depending on the sign of the hopping integrals, reflecting the
difference of band structure at the Fermi level.
\end{abstract}
\pacs{71.10.Hf, 71.27.+a, 71.10.-w}
\maketitle
%\begin{multicols}{2}
\narrowtext 
%%%%%%%%%%%%%%%%%%%%%%%%%%%%%%%%%%%%%%%%%%%%%%%%%%%%%%%%%%%%%%%%%%%%%%%%%%%%%%
%
{\it Introduction}---
For a long time, the exotic phenomena emergent in geometrical frustration
have been fascinating but challenging subjects of research in condensed
matter physics~\cite{Moessner-R}. An essential difficulty of frustrated
systems is attributed to their highly-degenerate ground states in the
classical sense. It is common practice to determine how
the degeneracy is removed or how the frustration is minimized by adding
quantum fluctuations.
%
%For example, quantum fluctuations yield a 120$^\circ$ structure for the
%ground state of an antiferromagnetic interaction on a triangle (the
%simplest case of an interaction frustrated due to lattice geometry).  In
%the context of the triangular $S=1/2$ antiferromagnet a spin liquid
%state was proposed by Anderson~\cite{Anderson}.
%
On a related issue, one of the most intensively studied systems is
quantum spins on the Kagom\'{e} lattice, which consists of
corner-sharing triangles. Very recently, so-called herbertsmithite has
been revealed as a promising candidate for an experimental realization
of an $S=1/2$ Kagom\'e antiferromagnet.  Both
experimentally~\cite{Helton,Mendels} and
theoretically~\cite{Sindzingre-L}, the low-temperature phase of this
system is characterized by a spin liquid state with a small or possibly 
even no energy gap.

When we turn our attention to itinerant systems on the Kagom\'e lattice,
a greater variety of physical phenomena can be observed. Of particular
interest is the competition of distinct ordered states associated with
the charge degrees of freedom. From this point of view, an intriguing
system is that of 1/3-filled Kagom\'{e} lattice fermions. If long-range
repulsive interactions are taken into account, there exists a huge
number of (nearly) degenerate charge-ordered (CO) states in the
strong-coupling regime; a melting of the CO may take place as some ratio
of the interaction and hopping.

\begin{figure}[b]
\begin{center}
\includegraphics[width=8.5cm]{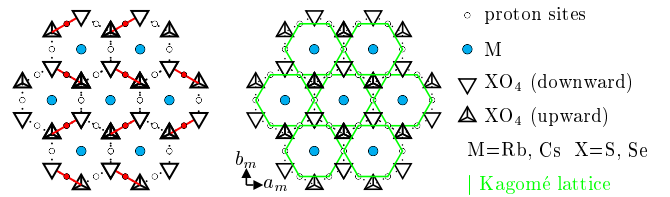}
\end{center}
\caption{(Color online) Lattice structure of hydrogen-bonded crystals
 M$_3$H(XO$_4$)$_2$ where the proton sites form the Kagom\'{e} 
 lattice (right figure). The left figure presents an example of protons order 
 below the critical temperature. The solid (red) lines denote hydrogen 
 bonds.
 }\label{fig:MHXO}
\end{figure}

An experimental candidate to realize this situation is hydrogen-bonded
crystals of alkali-hydrosulfates or hydroselenates M$_3$H(XO$_4$)$_2$
(M=Cs,Rb,K,Tl,\ X=S,Se)~\cite{Kamimura}. These materials exhibit an
abrupt change in the electrical conductivity accompanied with a
ferroelastic deformation around $T_{\rm c}=400$K: above $T_{\rm c}$ the
conductivity within the {\it a}-{\it b} plane is very high and it seems
to be related to the hopping motion of protons, i.e., super-proton
conductivity; below $T_{\rm c}$, the protons appear to be localized in
some ordering pattern which satisfies the condition that one tetrahedron
of XO$_4^-$ has only one hydrogen bond (see Fig.~\ref{fig:MHXO}).  In
the regime where only proton ordering is taken into account, an
effective model can provide a description of the protons forming the
Kagom\'e network, as discussed later in the paper. Since the number of
corner-sharing triangles is equivalent to the number of tetrahedra, the
critical behavior in these materials can be described by a
metal-insulator transition (MIT) in the Kagom\'e lattice at 1/3 filling.

Another possible experimental realization may be given by recently
developed laser-cooling techniques. It was
reported~\cite{Santos-B-C-E-F-M} that a two-dimensional optical {\it
trimerized} Kagom\'e lattice was constructed and that the coupling
constants could be controlled using a triple laser beam design.
Furthermore, a scheme proposed this year would construct an optical
Kagom\'e lattice at 1/3 filling with the use of just two standing waves
and more controllable interactions~\cite{Ruostekoski}.

Motivated by the optical lattice systems, a model of hard-core bosons on
the Kagom\'e lattice was studied with large scale quantum Monte Carlo
simulations and phenomenological dual vortex
theory~\cite{Isakov-W-M-S-K,Sengupta-I-K}. At 1/3 and 2/3 fillings a
weak first-order superfluid-solid transition induced by the
nearest-neighbor repulsion was found. Thus, in our work presented here
it is natural to consider how this transition occurs in fermion
systems. Although the effective model can be shown to be equivalent for
hard core bosons and spinless fermions in the strong-coupling limit,
features for each, in the weak to intermediate coupling regimes, may
differ completely due to the absence of a band picture in boson systems;
here the distinction between the two cases is non-trivial.

In this Letter, we therefore study the MIT of the spinless fermion model
on the Kagom\'e lattice at 1/3 filling with the nearest-neighbor
repulsive interactions using the exact-diagonalization (ED) and
density-matrix renormalization group (DMRG) techniques. Assuming a CO
state in the strong-coupling regime, we demonstrate that the nature of
the MIT strongly depends on the sign of hopping integral (inversion of
the sign corresponds to a switch between $1/3$ and $2/3$
fillings). Lastly, the random phase approximation (RPA) is employed to
confirm the numerical results.

\begin{figure}[t]
\begin{center}
\includegraphics[width=7.5cm]{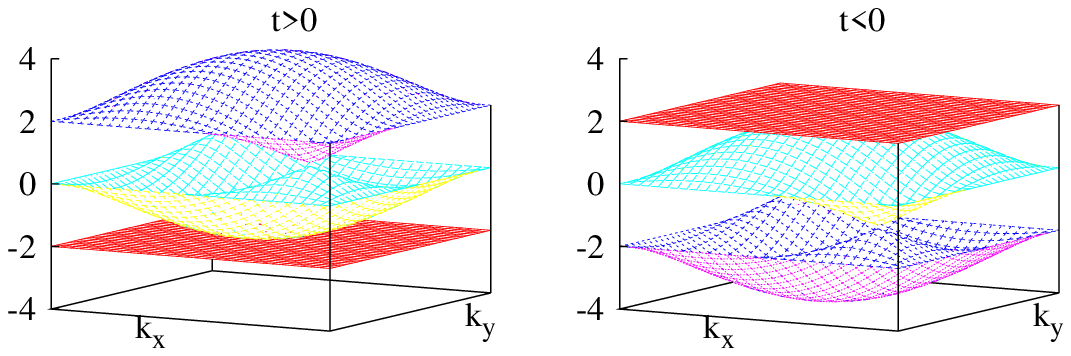}
\end{center}
\caption{(Color online) Band structure of the Kagom\'{e} lattice
fermions for positive (left) and negative (right) hopping energies.
}\label{fig:band}
\end{figure}

{\it Model}---
The Hamiltonian is
\begin{equation}
 {\cal H}=t\sum_{\langle i,j \rangle}
  (c^\dagger_i c^{\vphantom{\dagger}}_j + {\rm H.c.}) 
  +V\sum_{\langle i,j \rangle}n_i n_j,
  \label{tvham}
\end{equation}
where $c^{\dagger}_j$ ($c^{\vphantom{\dagger}}_j$) is a creation
(annihilation) operator of a spinless fermion and 
$n_j$ ($=c^\dagger_jc^{\vphantom{\dagger}}_j$) is the corresponding number operator. 
The repulsive interaction $V$($>0$) is assumed to act only between 
neighboring sites $\langle i,j \rangle$. In the non-interacting case $V=0$, 
the dispersion relation is given by three bands: 
\begin{align}
\nonumber
\varepsilon(\bm{k})&=-2t,\\
&t\left[1 \pm \sqrt{1+8\cos\frac{k_x}{2}
\cos\frac{k_y}{2}\cos\left(\frac{k_x-k_y}{2}\right)}\right]
  \label{fig:ek}
\end{align}
(see Fig.~\ref{fig:band}). It is particularly worth noting that 
the flat band at 1/3 filling is only filled when $t>0$. Since the density of 
states at the Fermi level diverges, the physical properties 
in the weak- to intermediate-coupling regime are expected to 
greatly differ from those for $t<0$. We note that our model 
with $t>0$ at $1/3$ filling is equivalent to 
that with $t<0$ at $2/3$ filling via the particle-hole 
transformation. 

\begin{figure}[b]
\begin{center}
\includegraphics[width=7.5cm]{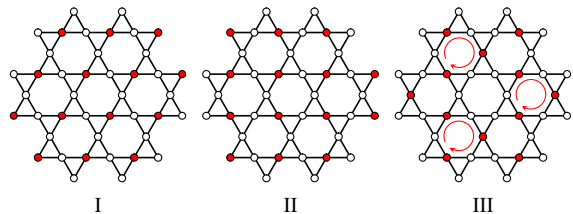}
\end{center}
\caption{(Color online) Charge-ordering patterns in the 1/3-filled 
 Kagom\'{e} lattice characterized by wave vectors I $(q_{\rm a},q_{\rm b})=(0,0)$, 
 II $(0,\pi)$, and III $(2\pi/3,2\pi/3)$. The circle on pattern III 
 corresponds to the cyclic hopping process by Eq.(\ref{eff_model}).
 }  \label{fig:CO_patterns}
\end{figure}

Let us now consider the case of $t=0$ with $V>0$. The nearest-neighbor
repulsions $V$ are minimized when each the corner-sharing triangle is
occupied by exactly one fermion.
%(and the two neighboring sites are never
%occupied)
%All other configurations have a larger potential energy.
This means
that the system is in macroscopically degenerate CO state. However, the
degeneracy may be lifted when a finite hopping is introduced, and then a
CO state with some periodicity must have the lowest energy. The
candidates of the CO patterns are quoted in
Fig.~\ref{fig:CO_patterns}. By looking at the pattern III, one realizes 
the kinetic energy can be gained through a cyclic
hopping process of three fermions on each hexagon. Thus, the lifting
takes place with processes of order $|t|^3/V^2$ ( lower order
contributions, e.g. ~$t^2/V$, are the same for all configurations and
therefore contribute a constant energy shift rather than lifting the
degeneracy).  The effective Hamiltonian is derived
as~\cite{Runge-F,Fulde-P,O'Brien-P-F}
\begin{equation}
 \mathcal{H}_{\rm eff}=\frac{12|t|^3}{V^2}\sum_{\rm hexagon}
  (c_1^{\dag}c_3^{\dag}c_5^{\dag}
  c_2^{\mathstrut}c_4^{\mathstrut}c_6^{\mathstrut}+{\rm H.c.}),
  \label{eff_model}
\end{equation}
where the sum is over all hexagons on the lattice and the six sites on 
each hexagon are labeled by 1-6 in a clockwise manner. This is equivalent 
to the quantum dimer model on the honeycomb lattice which has 
a plaquette-ordered ground state with three-fold degeneracy~\cite{Moessner-S-C}. 
The effective model would be at least qualitatively valid 
as long as the system is a CO insulator. We note that here the sign of 
hopping integral has no influence on the physics because the honeycomb 
lattice has a bipartite structure. 

\begin{figure}[b]
\begin{center}
\includegraphics[width=7.5cm]{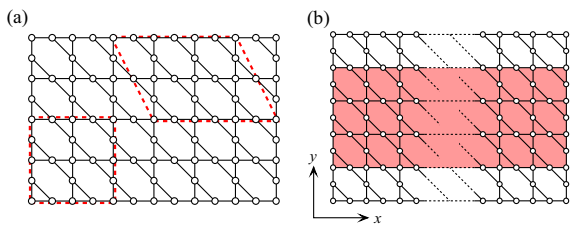}
\end{center}
\caption{(Color online) (a) Finite clusters of the Kagom\'{e} lattice
applied for the ED method. The 12- and 18-site clusters are used to
obtain $E_{\rm I}-E_{\rm II}$ and $E_{\rm II}-E_{\rm III}$,
respectively.  (b) $L_x \times 3$ cluster (shaded regime) for the DMRG
method, where the open and periodic boundary conditions are taken for
the $x$ and $y$ directions, respectively.  } \label{fig:geometries}
\end{figure}
\begin{figure}[t]
\begin{center}
\includegraphics[width=7.0cm]{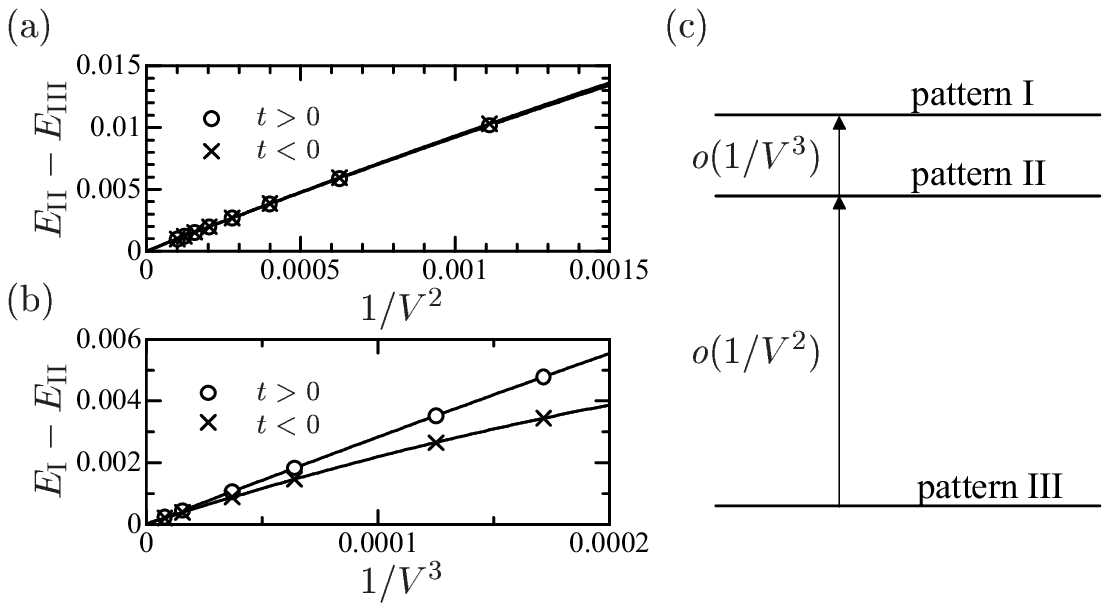}
\end{center}
\caption{(Color online) (a) Energy differences $E_{\rm II}-E_{\rm III}$ 
as a function of $1/V^2$ in units of $|t|$, and (c) $E_{\rm I}-E_{\rm II}$ as a function 
of $1/V^3$. (c) Relationship of the energies $E_{\rm I}$, $E_{\rm II}$, 
and $E_{\rm III}$ corresponding to the CO patterns I~-~III in Fig.~\ref{fig:CO_patterns}.
} \label{fig:GS_energy}
\end{figure}

%%%%%%%%%Aroon got this far-30 November 09%%%%%%%%%%
%%%%%%%%%%%%%%%%%%%%%%%%%%%%%%%%%%%%
{\it Numerical analysis}---
First, we compare the lowest energies of each CO pattern to determine
if pattern III is indeed stabilized for $V \gg |t| >0$. 
We apply the ED technique with periodic boundary conditions. 
A 12- (18-)site cluster is used to calculate an energy difference 
$E_{\rm I}-E_{\rm II}$ ($E_{\rm II}-E_{\rm III}$) [see
Fig.~\ref{fig:geometries}],
where $E_{\rm I}$, $E_{\rm II}$, and $E_{\rm III}$ are the
%ground-state
energies of the CO patterns I, II, and III, respectively. The
periodicity of wave functions is checked in order to identify one CO
state from another.  As shown in Fig.~\ref{fig:GS_energy}, for both
signs of $t$, $E_{\rm III}$ has the lowest ground-state energy while
$E_{\rm II}$ is the second-lowest in the large-$V$ regime; the energy
differences are scaled like $E_{\rm II}-E_{\rm III} \sim {\cal
O}(V^{-2})$ and $E_{\rm I}-E_{\rm II} \sim {\cal O}(V^{-3})$. Since the
second-order contributions ${\cal O}(|t|/V^2)$ cancel in the energy
differences, the behavior of $E_{\rm II}-E_{\rm III}$ seems to be
consistent with the existence of an energy gain ${\cal O}(|t|^3/V^2)$
from the ring exchange process in the effective Hamiltonian
(\ref{eff_model}).  As expected, $E_{\rm II}-E_{\rm III}$ has no
dependence on the sign of $t$ in the large-$V$ region as seen in
Fig.~\ref{fig:GS_energy} (b).

%As expected in the strongly correlated limit, the
%symmetry of the effective Hamiltonian means that $E_{\rm II}-E_{\rm
%III}$ has no dependence on the sign of $t$ (as seen in
%Fig.~\ref{fig:GS_energy} (b)).

\begin{figure}
\begin{center}
\includegraphics[width=8.0cm]{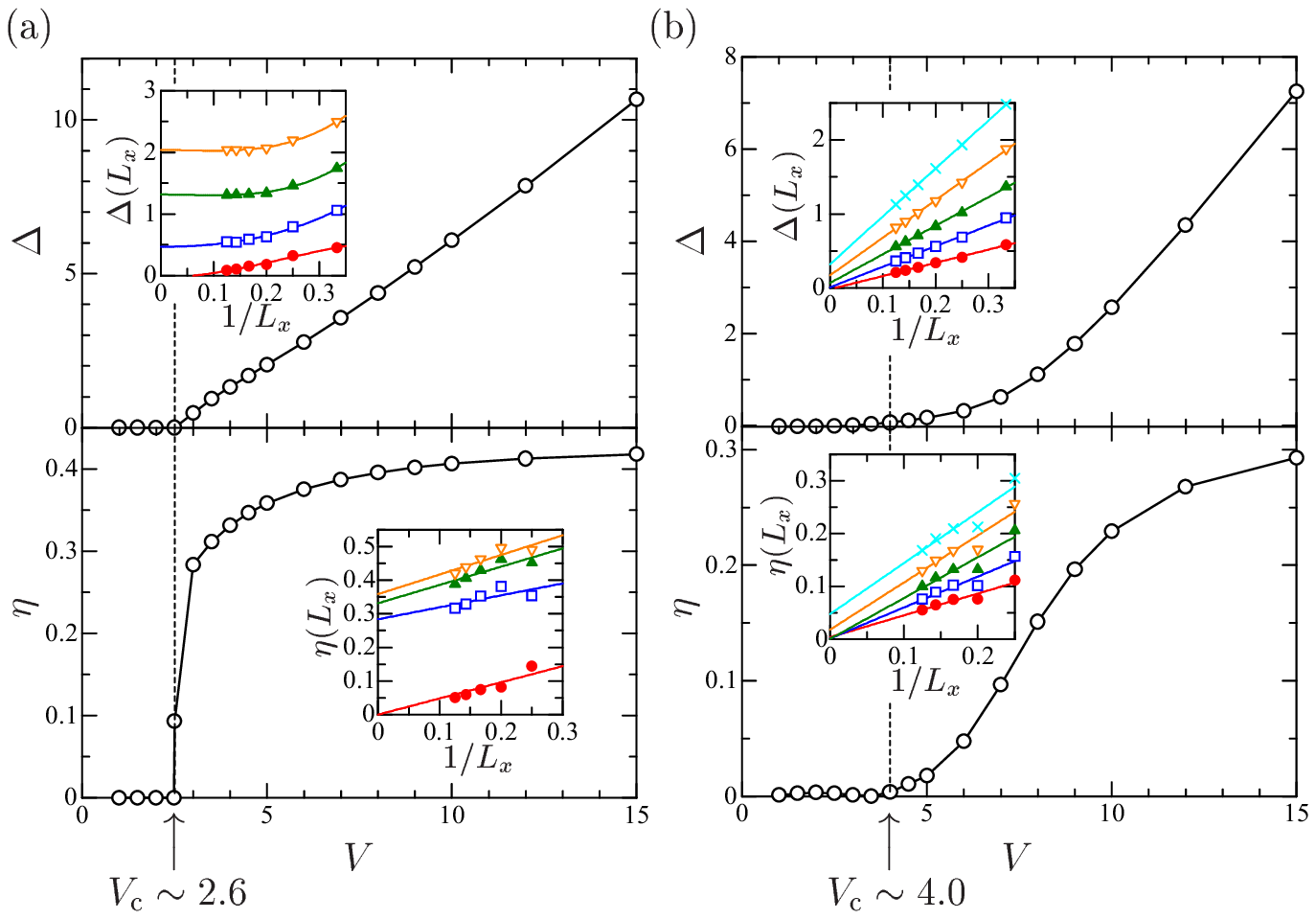}
\end{center}
\caption{(Color online) Single-particle gap $\Delta$ and order parameter 
$\eta$ as a function of the nearest-neighbor repulsion $V$ for (a) positive 
and (b) negative $t$ values. Insets: Finite-size scaling of each quantity 
for (a) $V=2, 3, 4$, and $5$; and (b) $V=2, 3, 4, 5$, and $6$, from bottom 
to top.
  }\label{fig:gap_order}
\end{figure}

Next, we show that the system indeed exhibits the CO-melting MIT 
at a finite critical interaction $V=V_{\rm c}$. 
To find the MIT critical point, the single-particle gap is calculated 
using the DMRG method. The DMRG calculation is performed on a $L_x \times L_y$ 
cluster shown in Fig.~\ref{fig:geometries}(b), where the periodic boundary 
conditions are imposed in the $y$ direction with three unit cells 
and the open boundary conditions in the $x$ direction with $L_x$ unit cells. 
We study several lengths of the cluster with up to $L_x=8$ ($78$ sites) 
and extrapolate the finite-size results to the thermodynamic limit 
$L_x \to \infty$. The single-particle gap is thus given by 
$\Delta=\lim_{L_x \to \infty}\Delta(L_x)$ with 
$\Delta(L_x)=E(N_f+1,L_x,L_y)+E(N_f-1,L_x,L_y)-2E(N_f,L_x,L_y)$, 
where $E(N_f,L_x,L_y)$ is the ground-state energy of the corresponding 
cluster with $N_f$ fermions. The upper panels of Figure~\ref{fig:gap_order} 
show the extrapolated values of $\Delta$ as a function of $V$ in units of $|t|$. 
The critical points are estimated as $V_{\rm c} \sim 2.6$ and $4.0$ for 
$t>0$ and $t<0$, respectively. The behaviors of the gap opening are 
quite different: the gap for $t>0$ increases almost linearly, while 
it rises gradually for $t<0$ like the Berezinskii-Kosterlitz-Thouless 
transition.

In order to gain further insight, we also calculate the order parameter 
corresponding to the CO pattern III. The CO phenomenon is observed 
as a state with a broken translational symmetry: actually, there 
are two or more degenerate ground states and {\it one} configuration 
of the degenerate states is picked out as the ground state by an initial 
condition of the DMRG calculation, as a consequence of the cylindrical boundary 
conditions~\cite{Nishimoto-H}. The order parameter is now defined as 
the amplitude of the charge-density modulation $\eta(L_x)= n_1-\frac{1}{2}(n_2+n_3)$ 
for a $L_x \times 3$ cluster, where $n_1$ and $n_2$ ($=n_3$) are 
the charge densities of fermion-rich and fermion-poor sites on a triangle 
at the center of the cluster. In the lower panels of Figure~\ref{fig:gap_order}, we show 
the extrapolated results of order parameter $\eta$ to the thermodynamic 
limit, i.e., $\eta=\lim_{L_x \to \infty} \eta(L_x)$, as a function of $V$. 
We find that the estimated critical points are almost the same as those 
obtained from $\Delta$ for both signs of $t$. As seen similarly for $\Delta$., the behavior of 
$\eta$ differs markedly depending on the sign of t. For $t>0$, the transition appears to be nearly discontinuous, 
which might be analogous to that in a melting MIT observed in hard-core 
bosons on the triangular lattice~\cite{Wessel-T}; for $t<0$, it looks continuous, which 
looks like a weakly first-order phase transition for hard-core bosons 
on the Kagom\'e lattice~\cite{Isakov-W-M-S-K}. Hence, we conclude that 
the criticality for positive $t$ is in stark contrast to that for negative $t$.

Now, we comment on the difference in values of the MIT critical points 
between $t>0$ and $t<0$. As mentioned above, all the triangles are only 
allowed to be singly occupied in the CO state. Therefore, our system 
can be mapped into a single-band half-filled Hubbard model on honeycomb 
lattice with on-site Coulomb interaction ${\cal O}(V)$ if we regard 
each triangle as a single site. The single-particle gap is then expected to 
behave like $\Delta \approx V-W_{\rm eff}$ in the large-$V$ region, 
where $W_{\rm eff}$ is the effective bandwidth. Actually, as seen in 
Fig.~\ref{fig:gap_order}, the gap behaves like $\Delta \sim V-5t$ and 
$\Delta \sim V-8|t|$ for $t>0$ and $t<0$, respectively. 
In general, the MIT critical strength is roughly scaled by 
the bandwidth, i.e., $V_{\rm c} \propto W_{\rm eff}$. The assumption 
of the effective bandwidths $W_{\rm eff}=5t$ for $t>0$ and $8|t|$ for 
$t<0$ are compatible with the obtained critical points. The narrower 
bandwidth for $t>0$ might be interpreted as a remnant of the flat band 
in the noninteracting case~\cite{Poilblanc-T}.

{\it Random phase approximation}---
In addition, the MIT is investigated analytically using the RPA~\cite{Aonuma-N}. 
The charge susceptibility is given by the Kubo formula as
\begin{eqnarray}
 X_{\alpha\gamma}(\bm{q},i\omega_l)=\frac{1}{N_{\rm uc}}\int_0^{\beta
 \hbar}d\tau e^{i\omega_l\tau}
\left\langle n_{\bm{q},\alpha}(\tau)n_{-\bm{q},\gamma}(0)\right\rangle
\label{charge_susceptibility}
\end{eqnarray}
where $\left\langle \cdots \right\rangle$ is the ensemble average,
$n_{\bm{q},\alpha}$ is a number operator in momentum space,
$\beta=k_{\rm B}T$ is inverse temperature, $\omega_l$ is the Matsubara
frequency of bosons, and $N_{\rm uc}$ is the number of unit cells. Then,
applying the RPA, the charge susceptibility is obtained in the following
matrix form,
\begin{equation}
 X(\bm{q}, \i \omega_l)
 =\left[1+ 2 v(\bm{q})X^{(0)}(\bm{q}, \i \omega_l))\right]^{-1}
 X^{(0)}(\bm{q}, \i \omega_l),
 \label{mat}
\end{equation}
%\begin{equation}
% X(\bm{q}, \i \omega_l)
% =
% X(\bm{q}, \i \omega_l)^{(0)}
% - X(\bm{q}, \i \omega_l)^{(0)}
% 2 v(\bm{q})
% X(\bm{q}, \i \omega_l),
% \label{mat}
%\end{equation}
where the matrix elements of $v(\bm{q})$ are given by
\begin{align}
 &v_{12}(\bm{q}) = V \cos (q_x/2),\quad
 v_{13}(\bm{q}) = V \cos (q_y/2)\\
 &v_{23}(\bm{q})
 = V \cos ((q_x - q_y)/2),
 \label{coefficients}
\end{align}
%\begin{align}
% &v_{12}(\bm{q}) = V \cos \left( \dfrac{q_x}{2} \right),\quad
% v_{13}(\bm{q}) = V \cos \left( \dfrac{q_y}{2} \right)\\
% &v_{23}(\bm{q})
% = V \cos \left( \dfrac{q_x - q_y}{2}\right),
% \label{coefficients}
%\end{align}
with the relation $v_{ji}=v_{ij}$ and $v_{ii}=0$.  The bare
susceptibility is defined as
\begin{align}
\lefteqn{X(\bm{q}, \i \omega_l)_{\alpha \gamma}^{(0)}\equiv}\nonumber\\
&\frac{1}{N_{\rm u}} \sum_{\bm k} \sum_{\mu \nu}
 u_{\bm{k}, \alpha \mu}^*
 u_{\bm{k} + \bm{q}, \alpha \nu}^{\mathstrut}
 u_{\bm{k} + \bm{q}, \gamma \nu}^*
 u_{\bm{k}, \gamma \mu}^{\mathstrut}
 \Gamma (\bm{k}; \bm{q})_{\omega_l}^{\mu \nu},
\end{align}
where
\begin{equation}
 \Gamma (\bm{k}; \bm{q})_{\omega_l}^{\mu \nu} 
  =
 \frac{f_{\bm{k} + \bm{q}, \nu} - f_{\bm{k}, \mu}}
 {\i \hbar \omega_l - \varepsilon_{\bm{k} + \bm{q}, \nu} +
 \varepsilon_{\bm{k}, \mu}},
\end{equation}
with $f$ being the Fermi distribution function. $u_{\bm{k}}$ is a
matrix which diagonalizes the kinetic part of eq.~(\ref{tvham}) in the
momentum space $\bm{k}$.

\begin{figure}[t]
\begin{center}
 \includegraphics[width=6.0cm]{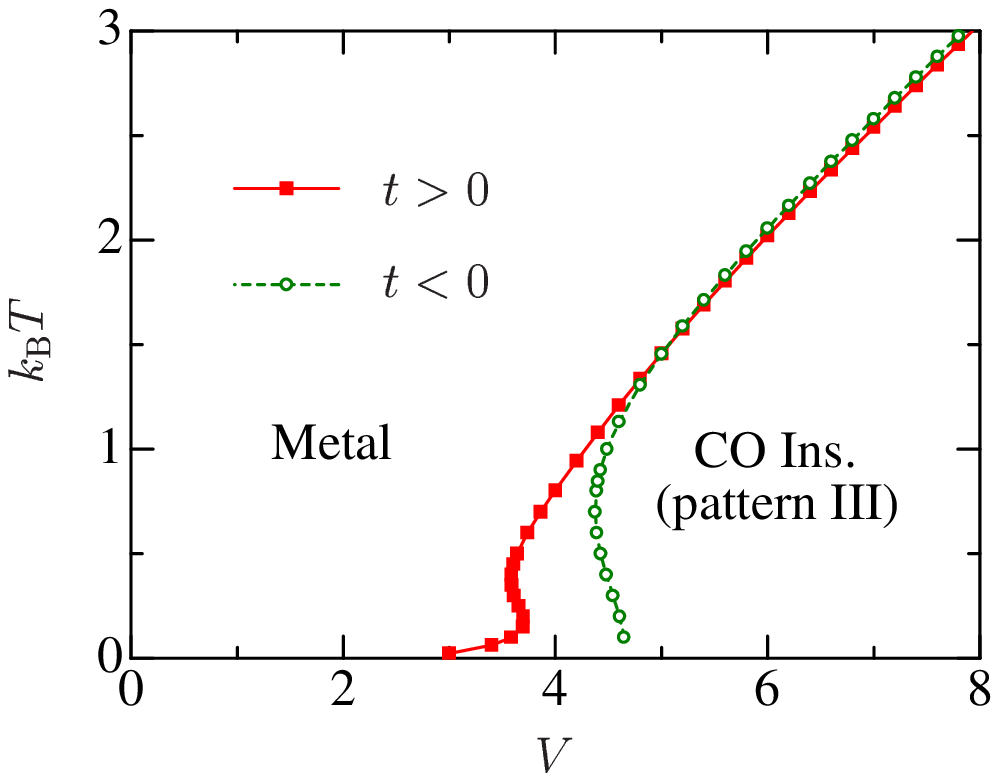}
\end{center}
\caption{(Color online) Phase diagram of the 1/3-filled $t$-$V$ model on
the Kagom\'{e} lattice obtained by the random phase approximation.}
\label{fig:PD}
\end{figure}

The MIT critical point at finite temperature is found by determining the
divergence of the diagonal part of eq.~(\ref{mat}) for the three sets of
CO patterns, I, II, and III.  We found that the CO pattern III is always
most stable, in agreement with the ED result. The obtained phase diagram
is shown in Fig.~\ref{fig:PD}.  At high temperature ($k_{\rm B}T/|t|\gg
1$), the MIT phase boundary is almost independent of the sign of $t$;
however, for decreasing temperature the boundaries gradually move apart
from each other.  The critical points are estimated as $V_{\rm c} \sim
3$ and $4.5$ for $t>0$ and $t<0$, respectively.  These values seem to be
not only qualitatively but also quantitatively consistent with those
obtained by the DMRG method. It is also interesting that the
boundaries show reentrant behaviors.

{\it Conclusion}--- We study the metal-insulator transition of the
spinless fermions on the Kagom\'{e} lattice at 1/3-filling using the
exact diagonalization, density-matrix renormalization group, and random
phase approximation techniques.  In the region of large $V$, the CO pattern 
in the ground-state is consistent with what is predicted by the ring exchange 
model. The behaviors of the single-particle 
gap opening for positive and negative hopping integrals are qualitatively quite different; 
this is further reflectedin the differing values of the critical point $V_c$.
These differences may arise from the distinct nature of the density of states at 
the Fermi level in each case. A possible further extension of this work 
is a consideration of spin degrees of freedom. This is of interest because 
spins in the CO phase are also frustrated and the configuration is nontrivial.

{\it Acknowledgment}---
We thank T.~Aonuma, G.~Fiete, Y.~Ohta, and F.~Pollmann for very useful
discussions.  M.~N. acknowledges the visitors program at the
Max-Planck-Institut f\"{u}r Physik komplexer Systeme, Global Center of
Excellence Program ``Nanoscience and Quantum Physics'' of the Tokyo
Institute of Technology by MEXT, and Industrial Technology Research
Grant Program in 2005-2008 from the New Energy and Industrial Technology
Development Organization (NEDO) of Japan.
%
%%%%%%%%%%%%%%%%%%%%%%%%%%%%%%%%%%%%%%%%%%%%%%%%%%%%%%%%%%%%%%%%%%%%%%
% References
%%%%%%%%%%%%%%%%%%%%%%%%%%%%%%%%%%%%%%%%%%%%%%%%%%%%%%%%%%%%%%%%%%%%%%

%%%%%%%%%%%%%%%%%%%%%%%%%%%%%%%%%%%%%%%%%%%%%%%%%%%%%%%%%%%%%%%%%%%%
%\end{multicols}
\end{document}